\documentclass[showpacs,preprintnumbers,amsmath,amssymb]{revtex4}
\usepackage{latexsym}
\usepackage{amsmath}
\usepackage{amssymb}
\usepackage{graphicx}
\usepackage{dcolumn}
\usepackage{bm}
\graphicspath{{./}{./figs/}}

\begin{document}

\title{An evolving network model with community structure}\thanks{JOURNAL OF PHYSICS A: MATHEMATICAL AND GENERAL 38 (2005) 9741-9749.}
\author{Chunguang Li$^{1,2}$}
\email{cgli@uestc.edu.cn}
\author{Philip K. Maini$^2$}
\affiliation{ $^1$ Centre for Nonlinear and Complex Systems,
School of Electronic Engineering, University of Electronic Science
and Technology of China, Chengdu, 610054, P. R. China.\\
$^2$ Centre for Mathematical Biology, Mathematical Institute,
University of Oxford, Oxford, OX1 3LB, United Kingdom.}
%\date{}
\begin{abstract}
Many social and biological networks consist of communities -
groups of nodes within which connections are dense, but between
which connections are sparser. Recently, there has been
considerable interest in designing algorithms for detecting
community structures in real-world complex networks. In this
paper, we propose an evolving network model which exhibits
community structure. The network model is based on the
inner-community preferential attachment and inter-community
preferential attachment mechanisms. The degree distributions of
this network model are analyzed based on a mean-field method.
Theoretical results and numerical simulations indicate that this
network model has community structure and scale-free properties.
\end{abstract}
\maketitle
\section{Introduction}
Complex networks are currently being studied across many fields of
science and engineering [1], stimulated by the fact that many
systems in nature can be described by models of complex networks.
A complex network is a large set of interconnected nodes, in which
a node is a fundamental unit usually with specific dynamical or
information content. Examples include the Internet, which is a
complex network of routers and computers connected by various
physical or wireless links; the World Wide Web, which is an
enormous virtual network of web sites connected by hyperlinks; and
various communication networks, food webs, biological neural
networks, electrical power grids, social and economic relations,
coauthorship and citation networks of scientists, cellular and
metabolic networks, etc. The ubiquity of various real and
artificial networks naturally motivates the current intensive
study of complex networks, on both theoretical and application
levels.

Many properties of complex networks have currently been reported
in the literature. Notably, it is found that many complex networks
show the small-world property [2], which implies that a network
has a high degree of clustering as in some regular network and a
small average distance between nodes as in a random network.
Another significant recent discovery is the observation that many
large-scale complex networks are scale-free. This means that the
degree distributions of these complex networks follow a power law
form $P(k)\sim k^{-\gamma}$ for large network size, where $P(k)$
is the probability that a node in the network is connected to $k$
other nodes and $\gamma$ is a positive real number determined by
the given network. Since power laws are free of characteristic
scale, such networks are called ``scale-free networks'' [3]. The
scale-free nature of many real-world networks can be generated by
a mechanism of growing with preferential attachment [3].

Communities are defined as collections of  nodes within which
connections are dense, but between which connections are sparser.
There are many real-world networks which exhibit community
structure, and community structures are supposed to play an
important role in many real networks. For example, communities in
a citation network might represent related papers on a single
topic [4]; communities on the web might represent pages on related
topics [5]; communities in a biochemical network or neuronal
system might correspond to functional units of some kinds [6, 7];
communities also play an important role in information networks
[8]. There have been several investigations into designing
algorithms for detecting community structure in large-scale
complex networks [9-10]. To study the effects of community
structure on network properties and dynamics, the modelling of
real networks with community structure is very important. However,
most of the existing evolving network models do not take the
community structure into account. In [11], a networked seceder
model was proposed to illustrate group formation in social
networks. In [12], a bipartite growing network model for social
community with group structures was proposed. In [13], a social
network model based on social distance attachment was proposed,
which can exhibit community structure. In [14], the authors
proposed a growing network model with community structure.
However, there is the possibility within the network model that a
node belongs to a community but has no connections with nodes in
this community but has connections with nodes in other
communities, which is unacceptable.

In this paper, we propose an evolving network model with community
structure based on the inner-community preferential attachment and
inter-community preferential attachment mechanics. We use a
mean-field method to analyze the degree distributions of this
network model. Numerical simulations are also performed to
investigate the properties of this network model. Some more
realistic generalizations and extensions are also discussed in the
the Conclusions and Remarks Section.

\section{Network Model}
In this section, we describe the growing mechanics of the proposed
network model. For simplicity, we consider only undirected network
models in this paper. We assume there is a total of $M\,(M\geq 2)$
communities in the network. The proposed model is defined by the
following scheme:

{\bf Step 1 - Initialization}:  Start from a small number $m_0\,\,
(m_0>1)$ of fully connected nodes in each community. Use
$\frac{M(M-1)}{2}$ inter-community links to connect each community
to the other $M-1$ communities, so that there is a link between
each community. The nodes to which the inter-community links
connect are selected randomly in each community. For example, Fig.
1 shows an initial network with $M=3$ and $m_0=3$.
\begin{figure}[htb]
\centering
\includegraphics[width=7cm]{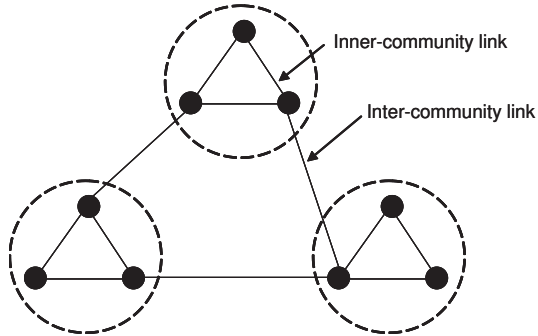}
\caption{An initial network example with $M=3, m_0=3$}
\end{figure}

{\bf Step 2 - Growth}: At each time step, a new node is added to a
randomly selected community. The new node will be connected to
$m\,\, (1\leq m\leq m_0)$ nodes in the same community through $m$
\emph{inner-community links} (defined as the links that connect
nodes in the same community), and with probability $\alpha$
connected to $n\,\, (1\leq n \leq m)$ nodes (none with probability
$1-\alpha$) in the other $M-1$ communities through
\emph{inter-community links} (defined as the links that connect
nodes among different communities).

{\bf Step 3 - Preferential Attachments}:

a) Inner-community preferential attachment: When choosing nodes in
the same community (we denote it as the $j$th community) to which
the new node connects through inner-community links, we assume
that the probability $\Pi$ that a new node will be connected to
node $i$ in community $j$ depends on the inner-degree $s_{ij}$
(defined as the number of inner-links connected to node $i$) of
that node, such that
\begin{equation}
\Pi(s_{ij})=\frac{s_{ij}}{\sum_k s_{kj}}.
\end{equation}

b) Inter-community preferential attachment: When choosing the
nodes in other communities to which the new node connects through
inter-community links, we assume that the probability $\Pi$ that a
new node will be connected to node $i$ in community $k (k\neq j)$
depends on the inter-degree $l_{ik}$ (defined as the number of
inter-links connected to the node), such that
\begin{equation}
\Pi(l_{ik})=\frac{l_{ik}}{\sum_{m,n,n\neq j} l_{m,n}}.
\end{equation}

The motivation for using the inner-community preferential
attachment mechanics is the same as that in many other growing
network models. There also exist inter-community preferential
attachment phenomena in some real networks with community
structure. For example, in scientific collaboration networks, a
multidisciplinary researcher is more likely to be willing to study
other unknown fields to him/her, and has stronger desire to
collaborate with researchers in other fields than other
single-topic researchers. In friendship networks, an individual
with many inter-community links is more likely to make friends
with different kinds of people, and has higher probability of
making new friends with people in other communities than other
people with less inter-community links. These are all
inter-community preferential attachment phenomena.

After $t$ time steps, this scheme generates a network with
$Mm_0+t$ nodes, and
$[Mm_0(m_0-1)+M(M-1)]/2+mt+\mbox{integer}(\alpha n t)$ links in
the sense of mathematical expectation. The parameters $\alpha$ and
$n$ control the ratio between inter- and inner-community links.

We performed a numerical simulation with a total of $N=90$ nodes,
$m_0=m=3, n=1$ and the probability $\alpha=0.1$, that is, at each
time step, we connect a new node with 3 nodes in a selected
community and with probability $\alpha=0.1$ connect it to a node
in another community. For the purposes of clarity, the generated
network is shown in Fig. 2, in which nodes in each community are
randomly placed. As we can see, the network generated by the
proposed scheme exhibits community structure. A detailed study of
the properties of this network model will be presented in the next
section.

\begin{figure*}[htb]
\centering
\includegraphics[width=12cm]{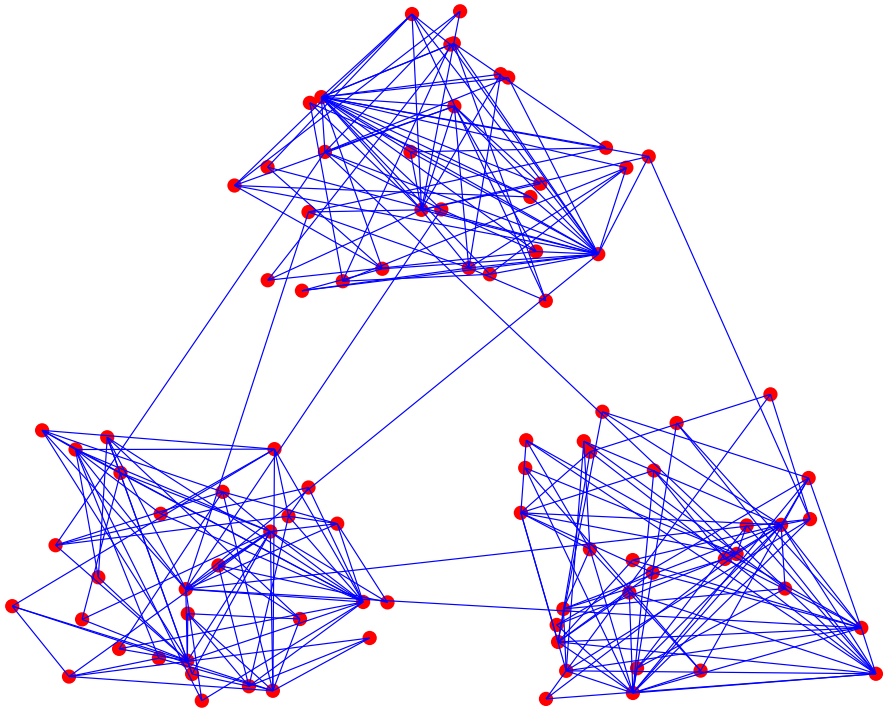}
\caption{A network generated by the proposed scheme: $N=90, M=3,
m_0=3, m=3, \alpha=0.1, n=1$ (the nodes in each community are
randomly placed).}
\end{figure*}

\section{Network Properties}
In this section, firstly we analytically calculate the degree
distribution $P(k)$, which is defined as the probability that a
randomly selected node has degree $k$ (a node has $k$
connections), and then we numerically study some properties of
this network model.

We use a mean-field method [15] to analyze the scaling property of
the network model. Firstly we analyze the inner-degree
distribution, and then extend the results to that of the
inter-degree and the total degree (defined as the sum of the
inner-degree and inter-degree). Similar to [15], we assume
$s_{ij}$ is continuous, and thus the probability
$\Pi(s_{ij})={s_{ij}}/ {\sum_k s_{kj}}$ can be interpreted as a
continuous rate of change of $s_{ij}$. Consequently, for a node
$i$ in community $j$, we have
\begin{equation}
\frac{\partial s_{ij}}{\partial t}=\frac{1}{M}m \frac{s_{ij}}
{\sum_k s_{kj}}
\end{equation}
and, noting that $\sum_k s_{kj}=2mt \frac{1}{M} +m_0(m_0-1)\approx
2mt \frac{1}{M}$ for large $t$ in the sense of mathematical
expectation, we have
\begin{equation}
\frac{\partial s_{ij}}{\partial t}\approx \frac{s_{ij}} {2t}\,.
\end{equation}

The solution of this equation, with the initial condition that
node $i$ in community $j$ was added to the system at time $t_i$
with connectivity $s_{ij}(t_i) = m$, is
\begin{equation}
s_{ij}(t)\approx m\left(\frac{t}{t_i}\right)^{0.5}.
\end{equation}
Using (5), the probability that a vertex has a connectivity
$s_{ij}(t)$ smaller than $k$ can be written as
\begin{equation}
P(s_{ij}(t)<k) = P(t_i> \frac{m^2t}{k^2}).
\end{equation}

Assuming that we add the nodes (including the initial nodes) at
equal time intervals to the network, the probability density of
$t_i$ is
\begin{equation}
P_i(t_i) = \frac{1}{Mm_0 + t}.
\end{equation}
By substituting (7) into (6), we have
\begin{equation}
P(t_i> \frac{m^2t}{k^2})=1- P(t_i\leq \frac{m^2t}{k^2})=1-
\frac{m^2t}{k^2(Mm_0+t)}.
\end{equation}

The probability density for $P(k)$ can be obtained using
\begin{equation}
P(k) = \frac{\partial P(s_{ij}(t)<k)}{\partial k} =
\frac{2m^2t}{Mm_0 + t}k^{-3},
\end{equation}
predicting that the inner-degree distribution obeys a power-law
distribution $P(k)\sim k^{-\gamma}$ with $\gamma=3$ independent of
$M$ and $m$.

Similarly,
\begin{equation}
\frac{\partial l_{ik}}{\partial t}=\frac{M-1}{M}\alpha
n\frac{l_{ik}}{\sum_{m,n,n\neq j} l_{m,n}}
\end{equation}
in which
\begin{equation}
\sum_{m,n,n\neq j} l_{m,n}=2 \frac{M-1}{M}\alpha nt
+[M(M-1)-(M-1)]
\end{equation}
in the sense of mathematical expectation. If $\alpha=0$, then no
inter-community links are added in the evolution of the network,
and the total number of inter-community links is always
$\frac{M(M-1)}{2}$. If $\alpha\neq 0$, the solution of this
equation, with the initial condition that node $i$ in community
$k$ was added to the network at time $t_j$ with inter-community
connectivity $l_{ik}(t_j) = \alpha n$ in the sense of mathematical
expectation, is
\begin{equation}
l_{ik}(t)=\alpha n\left(\frac{t+\beta}{t_j+\beta}\right)^{0.5}
\end{equation}
with $\beta=\frac{[M(M-1)-(M-1)]M}{2\alpha n(M-1)}$. If $\alpha$
is not so small, such that $2 \frac{M-1}{M}\alpha nt \gg
[M(M-1)-(M-1)]$ for large $t$ (usually, this is the case), then
\begin{equation}
\frac{\partial l_{ik}}{\partial t}\approx\frac{l_{ik}} {2t}.
\end{equation}

The solution of this equation, with the initial condition that
node $i$ in community $k$ was added to the network at time $t_j$
with inter-community connectivity $l_{ik}(t_j) = \alpha n$ in the
sense of mathematical expectation, is
\begin{equation}
l_{ik}(t)\approx \alpha n\left(\frac{t}{t_j}\right)^{0.5}.
\end{equation}
Similar to the above analysis of the inner-degree distribution,
the inter-degree distribution $P(k)$ can be written as
\begin{equation}
P(k)  = \frac{2(\alpha n)^2t}{Mm_0 + t}k^{-3},
\end{equation}
predicting that the inter-degree distribution also obeys a
power-law distribution $P(k)\sim k^{-\gamma}$ with $\gamma=3$
independent of $M, \alpha$ and $n$.

In this case, the total degree of node $i$ in community $j$ is
\begin{equation}
k_{ij}(t)=s_{ij}(t)+l_{ij}(t)\approx \left(m+\alpha
n\right)\left(\frac{t}{t_i}\right)^{0.5}.
\end{equation}
It is easy to show that the total degree distribution is
\begin{equation}
P(k)  = \frac{2(m+\alpha n)^2t}{Mm_0 + t}k^{-3}
\end{equation}
which indicates that the total degree also obeys a power-law
distribution $P(k)\sim k^{-\gamma}$ with $\gamma=3$.

Because $\alpha n<m$ (usually $\alpha n\ll m$), from (5) and (14)
(or (12)) we know that the inter-degrees of network nodes are
smaller than the inner-degrees of network nodes in the sense of
statistics. So, networks generated by this model will have
community structure.

Next, we numerically study the properties of the network model. We
consider a network generated by the proposed scheme with $M=3,
m_0=3, m=3, \alpha=0.3, n=1$ and $N=3000$. In the figures 3-5, the
slopes of the lines were obtained by cumulative distributions of
10 runs. The inner-degree distribution of this network is shown in
Fig. 3. As we can see, the inner-degree obeys a power-law
distribution $P(k)\sim k^{-\gamma}$ with $\gamma\approx 3$ (the
slope of the line is -3). The inter-degree distribution is shown
in Fig. 4, and the inner-degree obeys a power-law distribution
$P(k)\sim k^{-\gamma}$ with $\gamma\approx 2.9$ (the slope of the
line is -2.9). The total degree distribution is shown in Fig. 5,
and we observe that the total degree also obeys a power-law
distribution $P(k)\sim k^{-\gamma}$ with $\gamma\approx 3$. So,
all these three distributions are of power-law type, indicating
that the generated network is a scale-free network, and these
distributions verify the above analytical results. In Fig. 6 we
plot the time evolution of the total degree of two nodes, in which
one is an initial node and the other one is added to the network
at $t=60$. This is in good agreement with (16).
\begin{figure}[htb]
\centering
\includegraphics[width=7cm]{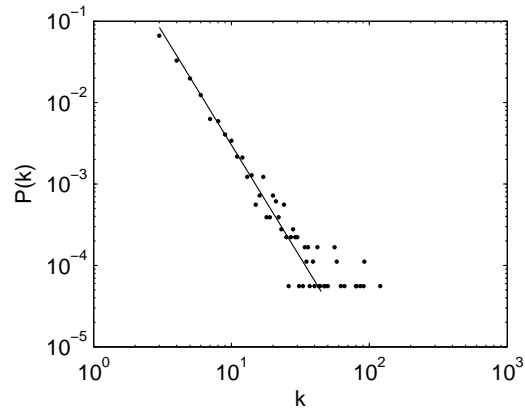}
\caption{Inner-degree distribution of a network with $N=3000, M=3,
m_0=3, m=3, \alpha=0.3, n=1$. As can be seen, the distribution is
well approximated by a straight line with gradient $-3$.}
\end{figure}
\begin{figure}[htb]
\centering
\includegraphics[width=7cm]{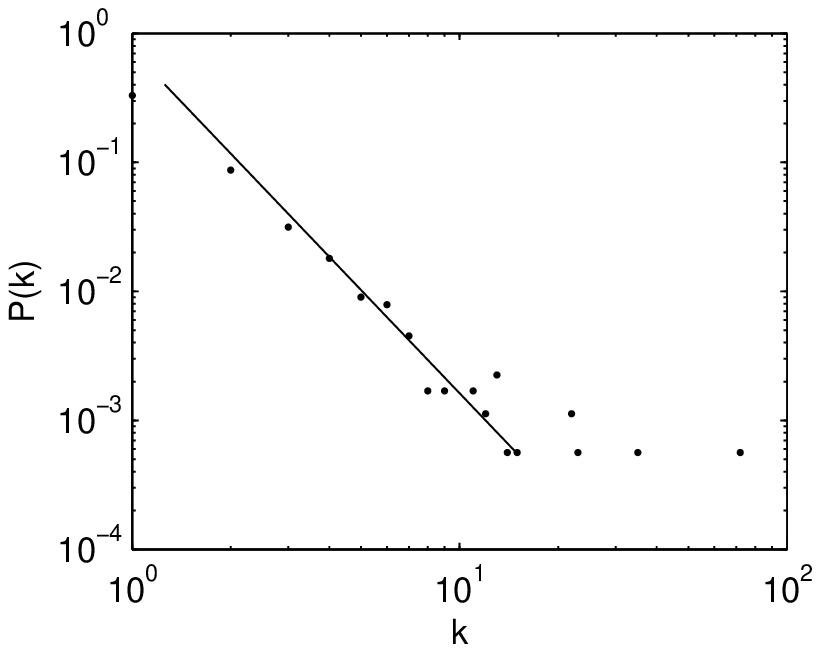}
\caption{Inter-degree distribution of a network with $N=3000, M=3,
m_0=3, m=3, \alpha=0.3, n=1$. As can be seen, the distribution is
well approximated by a straight line with gradient $-2.9$.}
\end{figure}
\begin{figure}[htb]
\centering
\includegraphics[width=7cm]{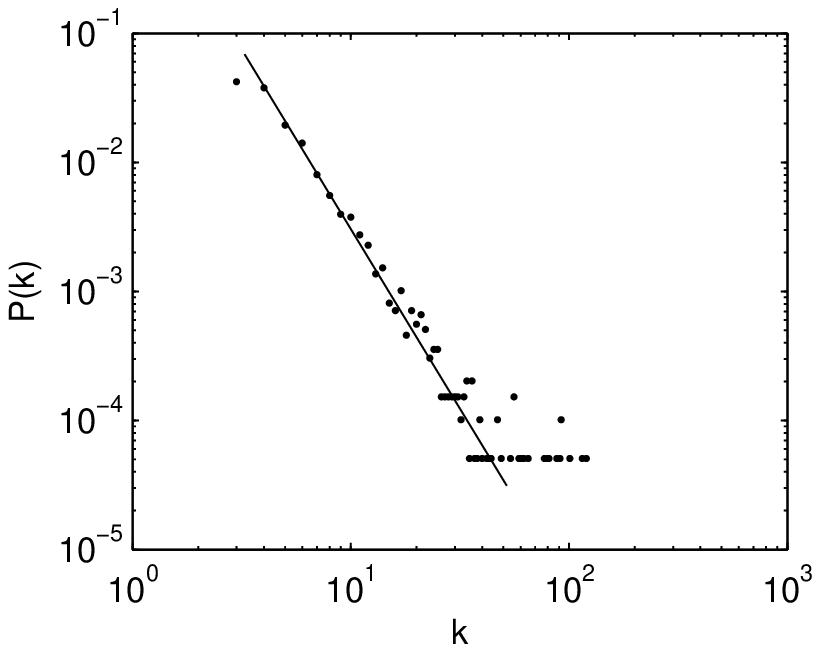}
\caption{Total degree distribution of a network with $N=3000, M=3,
m_0=3, m=3, \alpha=0.3, n=1$. As can be seen, the distribution is
well approximated by a straight line with gradient $-3$.}
\end{figure}
\begin{figure}[htb]
\centering
\includegraphics[width=7cm]{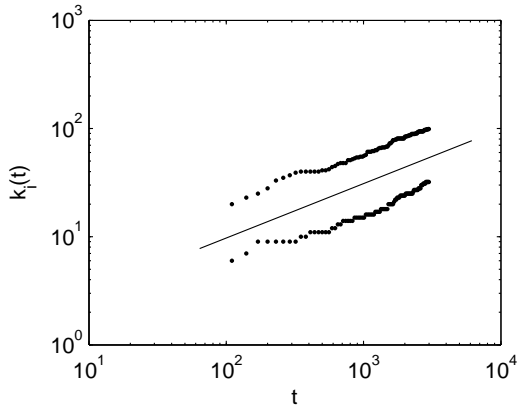}
\caption{Time-evolution of the degree of two nodes, one is an
initial node (the upper one) and the other one was added to the
network at $t=60$ (lower one). As can be seen, the time-evolution
of the degrees are well approximated by a straight line with
gradient $0.5$. }
\end{figure}

We also calculated the diameter $D$, the average path length $l$
and the clustering coefficient $C$ for this network. The detailed
definitions of these parameters can be found in [1]. The values of
these parameters are $D=7$, $l=4.0274$ and $C=0.0270$,
respectively. So the network has small average path length and
relatively large clustering coefficient (compared to a random
graph).

\section{Conclusions and remarks}
In this paper we proposed an evolving network model with community
structure. We have theoretically analyzed the scaling properties
of the network by using a mean-field approach. The analytical and
numerical results indicate that the network can produce community
structure, and the inner-degree, inter-degree and total degree all
obey power-law distributions, so the network has scale-free
properties. The network also has small average path length and
relatively large clustering coefficient (compared to a random
graph). Note that in [16] the authors proposed a growing network
model of two coupled networks, which idea is a little similar to
that in the present paper, but the authors didn't consider
community structure.

To make our model more understandable and to avoid unnecessarily
complicated notations, we made some simplifications in our model.
Some more realistic cases and extensions can be considered with
minor modification of our model, for example:
\begin{enumerate}
\item We didn't consider the formation of new communities during
the evolution of the network model. In some real world networks,
some new communities may appear during the evolution. With minor
modification, our model could allow from time to time the
introduction of new communities: if a node $i$ is the first of a
new community, one could only set links between $i$ and the old
communities at the time step of its introduction, and follow the
same evolving rules in the proceeding time steps as defined in
Section 2. It is easy to show that, with this modification, the
main results will not change.

\item In the preferential attachment mechanism, if a node $i$ is
not connected to a node $j$ (in the same community or different
communities) at the beginning of its introduction, it will never
be. In considering the resulting networks, this is of no problem,
but in considering the evolution process, it is somewhat
unrealistic. WE can overcome this problem, as follows: in each
time step, besides the Steps mentioned in Section 2, with
probability $\beta$, an existing node is selected, and we perform
the same inner- and inter- community attachments of this node to
all the other existing nodes that have no links with it. This will
surely not change the community structure of the network model.
\end{enumerate}

Future extensions of this work include considering other kinds of
inner- and inter- linking mechanisms other than preferential
attachment, such as preferential linking [17] and distance
preference [18]. Future extensions also include the modelling of
directed and weighted network models with community structure
[19], because many real-world networks with community structures
are directed and/or weighted [20]. Based on this network model, we
can study the effects of community structure on network dynamics,
such as the stability, synchronization, disease and rumor
spreading, and robustness. We can also introduce different types
of dynamical node models in different communities to study the
dynamical behavior of some real networks, for example biochemical
networks and brain networks.

\section*{Acknowledgement}
The authors are grateful to the anonymous reviewers for their
valuable suggestions and comments, which have led to the
improvement of this paper. The authors would like to thank Mrs Wei
Huang and Yigong Xiao for help in calculating some of the
statistical quantities. This research was supported by the Key
Program projects of the National Natural Science Foundation of
China under Grant No. 70431002, and the Youth Science and
Technology Foundation of UESTC under Grant L08010201JX04011.

\end{document}